\documentclass[twocolumn,reprint,aps,prl,showkeys,groupaddress,superscriptaddress,sectionbib,square,longbibliography]{revtex4-1}
\setcitestyle{numbers,square}
\usepackage{epsfig}
\usepackage{color}
\usepackage{times}
\usepackage{mathtools}
\usepackage{amsmath}
\usepackage{amssymb}
\usepackage{indentfirst}
\usepackage{graphicx}
\usepackage{bm}
\usepackage{longtable}
\usepackage{wrapfig}
\usepackage{comment}

\newcommand{\bey}{\begin{eqnarray}}
\newcommand{\eey}{\end{eqnarray}}
\newcommand{\vep}{\varepsilon}

\newcommand{\sg}{\sigma}
\newcommand{\bec}{\begin{center}}
\newcommand{\eec}{\end{center}}

\begin{document}
\title{Fifth-degree elastic potential for predictive stress-strain relations and elastic instabilities under large strain and complex loading in Si}

\author{Hao Chen}\email{chenhao.hitgucas@gmail.com}
\affiliation{Key Laboratory of Pressure Systems and Safety, Ministry of Education, School of Mechanical and Power Engineering, East China University of Science and Technology, Shanghai 200237, China}
\affiliation{Department of Aerospace Engineering, Iowa State University, Ames, Iowa 50011, USA}

\author{Nikolai~A. Zarkevich}\email{zarkev@ameslab.gov}
\affiliation{Ames Laboratory, U.S. Department of Energy, Iowa State University, Ames, Iowa 50011-3020, USA}

\author{Valery~I. Levitas}\email{vlevitas@iastate.edu}
\affiliation{Department of Aerospace Engineering, Iowa State University, Ames, Iowa 50011, USA}
\affiliation{Ames Laboratory, U.S. Department of Energy, Iowa State University, Ames, Iowa 50011-3020, USA}
\affiliation{Department of Mechanical Engineering, Iowa State University, Ames, Iowa 50011, USA}

\author{Duane~D. Johnson}\email{ddj@iastate.edu, ddj@ameslab.gov}
\affiliation{Ames Laboratory, U.S. Department of Energy, Iowa State University, Ames, Iowa 50011-3020, USA}
\affiliation{Department of Materials Science \& Engineering, Iowa State University, Ames, Iowa 50011, USA}

\author{Xiancheng Zhang}\email{xczhang@ecust.edu.cn}
\affiliation{Key Laboratory of Pressure Systems and Safety, Ministry of Education, School of Mechanical and Power Engineering, East China University of Science and Technology, Shanghai 200237, China}

\date{\today}

\begin{abstract}
Materials under complex loading develop large strains and often transition via an elastic instability, as observed in both simple and complex systems.
Here, we represent Si I under large strain in terms of Lagrangian strains by an 5$^{th}$-order elastic potential found by minimizing error relative to density functional theory (DFT) results.
The Cauchy stress -- Lagrangian strain curves for arbitrary complex loadings are in excellent correspondence with DFT results, including the elastic instability driving the Si {I}$\rightarrow${II} phase transformation (PT)  and the shear instabilities. PT conditions for Si {I}$\rightarrow${II} under action of cubic axial stresses are linear in Cauchy stresses in agreement with DFT predictions. Such elastic potential permits study of elastic instabilities and orientational dependence leading to different PTs, slip, twinning, or fracture, providing a  fundamental basis for continuum simulations of crystal behavior under extreme loading.

\end{abstract}

\keywords{Elastic potentials of fifth-degree; Elastic instability; Complex loading; Density functional theory; Large strains}
\maketitle

\section{\label{INTRODUCTION}Introduction}
{\par} Nonlinear, anisotropic elastic properties of single crystals determine material response to extreme loading, e.g., in shock waves, under high static pressure, and in defect-free crystals and nanoregions.   Elastic nonlinearity ultimately results in elastic lattice instabilities \cite{04,Grimvall-etal-2012,de2017ideal,Pokluda-etal-2015,Telyatniketal-16,wang1993crystal}.
Such instabilities dictate various phenomena, including phase transitions (PT, i.e., crystal-crystal \cite{mizushima1994ideal,Levitasetal-Instab-17,Levitasetal-PRL-17,zarkevich2018lattice}, amorphization \cite{binggeli1992elastic,kingma1993microstructural,brazhkin1996lattice,zhao2018shock,chen2019amorphization}, and melting \cite{tallon1989hierarchy,levitas2012virtual}), slip, twinning, and fracture, in particular,  theoretical strength in tension, compression, or shear \cite{de2017ideal,Pokluda-etal-2015,Umeno-Cerny-PRB-08,Cernyetal-JPCM-13,Telyatniketal-16,tang1994lattice}.  In addition, nonlinear elastic properties are necessary for simulations of material behavior under extreme static \cite{levitas2019tensorial}
or  dynamic \cite{clayton2014analysis,clayton2015crystal} loadings and near interfaces with significant lattice mismatch.

{\par}Notably, third-order \cite{zhao2007first,lopuszynski2007ab,Cao-etal-PRL-18} and seldom fourth-order elastic constants \cite{wang2009ab,telichko2017diamond}  are known for different crystals, as determined at small strains (e.g., 0.02-0.03). As such,  fourth-order elastic constants "should be treated as an estimation only," e.g., for Si \cite{telichko2017diamond}. Indeed, these elastic constants are not consistent with the observed equation of state of diamond \cite{levitas2019tensorial}.
Extrapolation to large strain is unreliable to describe the lattice instability (e.g., at 0.2 for Si \cite{zarkevich2018lattice} or 0.3-0.4 for B$_4C$ \cite{guo2019transgranular,an2014atomistic}).
Thus, to describe correctly elasticity, including any lattice instability, higher-order elastic potentials are required, and must be calibrated for a range of strain including lattice instability. For some loadings, stress-strain curves at finite strains are obtained \cite{Pokluda-etal-2015,Umeno-Cerny-PRB-08,Cernyetal-JPCM-13,Telyatniketal-16,zarkevich2018lattice,gao2019shear,guo2019transgranular,an2014atomistic}, yet this is insufficient for simulation of material behavior or describing lattice instabilities under arbitrary complex or extreme loadings.

{\par} Here, an elastic potential of fifth-degree for Si I under large strain was determined in terms of Lagrangian strains (all 6 components) by minimizing error with respect to density functional theory (DFT) results within large strain ranges that includes instability points. The Cauchy stress -- Lagrangian strain curves  for multiple complex loadings are in excellent agreement with DFT results, including elastic instability that drives the phase transformation to Si II and shear instabilities. Conditions for Si I$\rightarrow$Si II PT under action of cubic axial stresses are found to be linear in Cauchy stresses, as predicted by DFT. Importantly, lower-order potentials cannot yield similar precision in the description of stress-strain curves and elastic instabilities. Obtained elastic  potential opens possibility to study all elastic instabilities leading to different PTs, fracture, slip, and twinning, and represents a fundamental basis for continuum simulations of crystal behavior
under extreme static and dynamic loading including the above processes and their orientational dependence.

{\par} Due to the technological import, the deformation and PT properties of silicon have been studied intensely. The third-order elastic constants were found with DFT \cite{zhao2007first,lopuszynski2007ab} and   experiments \cite{hall1967electronic,mcskimin1964measurement}; however,
higher-order elastic constants were not reported. The lattice instability under two-parametric loadings was studied in \cite{Pokluda-etal-2015,Umeno-Cerny-PRB-08,Cernyetal-JPCM-13,Telyatniketal-16}.  Lattice instability conditions driving the  Si I$\rightarrow$II PT under action of the Cauchy stress tensor (6 independent stresses) were obtained in  \cite{Levitasetal-Instab-17,zarkevich2018lattice}, utilizing predictions from the phase field approach \cite{Levitas-IJP-13}.

{\par} {\it Nonlinear elastic potential.} Motion of an elastic body is described by vector function $x_i(X_j, t)$, where $t$ is time and $x_i$ (deformed) and $X_j$ (undeformed reference state) are the Cartesian coordinates of the position vector. The deformation gradient and finite Lagrangian strain are then $F_{ij} = {\partial x_{i}}/{\partial X_j}$ and  $\eta_{ij} = \frac{1}{2}(F_{ki}F_{kj}-\delta_{ij})$, respectively,  where $\delta_{ij}$ is the Kronecker delta (unit tensor) and Einstein summation notation is assumed.
Using Voigt notation to simplify presentation, i.e.,
$\eta_{ii}\rightarrow\eta_{i}$ (for ${i}$=1,2,3), and
$\eta_{23}\rightarrow\eta_{4}/2$, $\eta_{31}\rightarrow\eta_{5}/2$ and $\eta_{12}\rightarrow\eta_{6}/2$,
the specific internal energy per unit undeformed volume is, as a power-series expansion:
\begin{equation} \label{eq:5s}
\small{
\begin{aligned}
u =& u_0 + \frac{1}{2}c_{ij}\eta_{i}\eta_{j}+
\frac{1}{6}c_{ijk}\eta_{i}\eta_{j}\eta_{k}  \\
& + \frac{1}{24}c_{ijkl}\eta_{i}\eta_{j}\eta_{k}\eta_{l} +
\frac{1}{120}c_{ijkls}\eta_{i}\eta_{j}\eta_{k}\eta_{l}\eta_{s} + \cdots    ,
\end{aligned}
}
\end{equation}
where the $c_{...}$ are elastic moduli of  second, third, fourth, fifth and higher order. For crystals with cubic symmetry, Eq. (\ref{eq:5s}) is specified in supplemental material \cite{supplement} in cubic axes, with 3 second-,  6 third-, 11 fourth-,  and 18 fifth-order moduli \cite{teodosiu2013elastic}, found here using DFT. The second Piola-Kirchhoff (PK2) stress and the true Cauchy stress are defined as
\begin{equation} \label{eq:6}
    S_{i} = \frac{\partial u}{\partial \eta_{i}}; \quad
    \sg_{ij} =J^{-1} F_{ik} S_{km} F_{jm};    \quad  J=det F_{ik}.
\end{equation}

{\par}We performed DFT simulations supplementing our simulations in \cite{zarkevich2018lattice}, especially for shear strains and complex combined compression-shear loadings, and data are in \cite{supplement}.   Parameter identification procedure is carried out and results are presented in the natural  cubic  coordinate system.

{\par} {\it Fitting procedure.} Rather than determine certain set of elastic moduli from the distinct deformations \cite{mosyagin2017ab,wang2009ab},  we find all elastic moduli from second- to fifth-order by the least-squares regression using all of the DFT data we have (see supplemental for all DFT data used).
The error  $Z$ is a weighted sum of two terms related to the energy and PK2 stresses:
\begin{equation} \label{eq:7}
Z = \frac{1}{6M}\sum_{k=1}^{M}\sum_{i=1}^{6}\left | S_{i}^k-S_{i}^{k0} \right |^2 +\frac{1}{M} \sum_{k=1}^{M}w\left | u_k - u_k^0\right |^2.
\end{equation}
Here parameters without superscript $0$ designate results from approximate Eqs. (\ref{eq:5s})  and (\ref{eq:6}) and those with superscript  $0$ are DFT;
$M$ is the number of sets of results of DFT simulations,  and $w$ is the  weight factor.

{\par} {\it Fitted elastic moduli} are listed in Tables \ref{table:1} and \ref{table:2}, with comparison to the third-order elastic potential from other DFT results \cite{zhao2007first} and experiments \cite{hall1967electronic,mcskimin1964measurement}. 
The fourth- and fifth-order elastic moduli have no corresponding parameters from experiments and calculations to compare with.
In spite of some deviations (e.g., $c_{11}$, $c_{12}$, and $c_{123}$), the elastic constants are in good overall agreement with the previous DFT and experimental results.
As our main focus is on large strain and an elastic instability, we tolerate small discrepancies for small strains, and do not attempt to better fit second- and third-order elastic constant as then stress-strain curves from the elastic potential and DFT will be worse for large strain and the six-order potential will be required.

\begin{table} [h!]
   \begin{center}
\caption{Second- and third-order elastic constants for Si (in GPa), with comparison to other calculations and experiments.}
       \begin{tabular}{|l|l|l|l|l|}
\hline
            & Present Work & Other Theory{\cite{zhao2007first}} & Expt. 1{\cite{hall1967electronic}}       & Expt.2{\cite{mcskimin1964measurement}}       \\ \hline
$c_{11}$    & 151.76       & 162.07             & 165.04      & 165.77      \\ \hline
$c_{12}$    & 59.207       & 63.51              & 63.94       & 63.92       \\ \hline
$c_{44}$    & 77.90        & 77.26              & 79.51       & 79.62       \\ \hline
$c_{112}$   & -455.48      & -422               & -445$\pm$10 & -451$\pm$5  \\ \hline
$c_{111}$   & -653.38      & -810               & -795$\pm$10 & -825$\pm$10 \\ \hline
$c_{123}$   & -95.54       & -61                & -75$\pm$5   & -64$\pm$10  \\ \hline
$c_{144}$   & 22.56        & 31                 & 15$\pm$5    & 12$\pm$25   \\ \hline
$c_{155}$   & -304.11      & -293               & -310$\pm$5  & -310$\pm$10 \\ \hline
$c_{456}$   & -6.55        &                    &             &             \\ \hline
         \end{tabular}
\label{table:1}
     \end{center}
\end{table}

\begin{table} [h!]
    \begin{center}
\caption{Fourth- and fifth-order elastic constants for Si (in GPa).}
      \begin{tabular}{|l|l|l|l|l|l|}
\hline
$c_{1111}$  & 612.74       &    $c_{1112}$  & 2400.94   &   $c_{1122}$  & 1275.11   \\ \hline
$c_{1123}$  & 1053.03      &    $c_{1144}$  & 5070.79   &   $c_{1155}$  & 4049.80   \\ \hline
$c_{1255}$  & -2728.12     &    $c_{1266}$  & -513.56   &   $c_{1456}$  &  65.5     \\ \hline
$c_{4455}$  &  -576.86     &    $c_{4444}$  &  -2553.1  &   $c_{11111}$ &  465.42   \\ \hline
$c_{11112}$ &  -4330.81    &    $c_{11122}$ &  -3442.42 &   $c_{11123}$ &  -3765.50 \\ \hline
$c_{11155}$ &  -135641.41  &    $c_{11144}$ &  -225996.33 & $c_{11266}$ &  213.65   \\ \hline
$c_{12244}$ &  58582.68    &    $c_{11244}$ &  -10255.85  & $c_{11223}$ &  -1337.79 \\ \hline
$c_{11456}$ &  1063        &    $c_{12344}$ &  -5924.05 &    $c_{12456}$ &  -1653   \\ \hline
$c_{14444}$ &  20180.5     &    $c_{14455}$ &  43158.06 &    $c_{15555}$ &  32386.17 \\ \hline
$c_{15566}$ &  -83526.15   &    $c_{44456}$ &  625.51    &             &             \\ \hline
     \end{tabular}
\label{table:2}
   \end{center}
\end{table}

\begin{figure} [!tbp]
 \centering
 \includegraphics[width=0.4\textwidth]{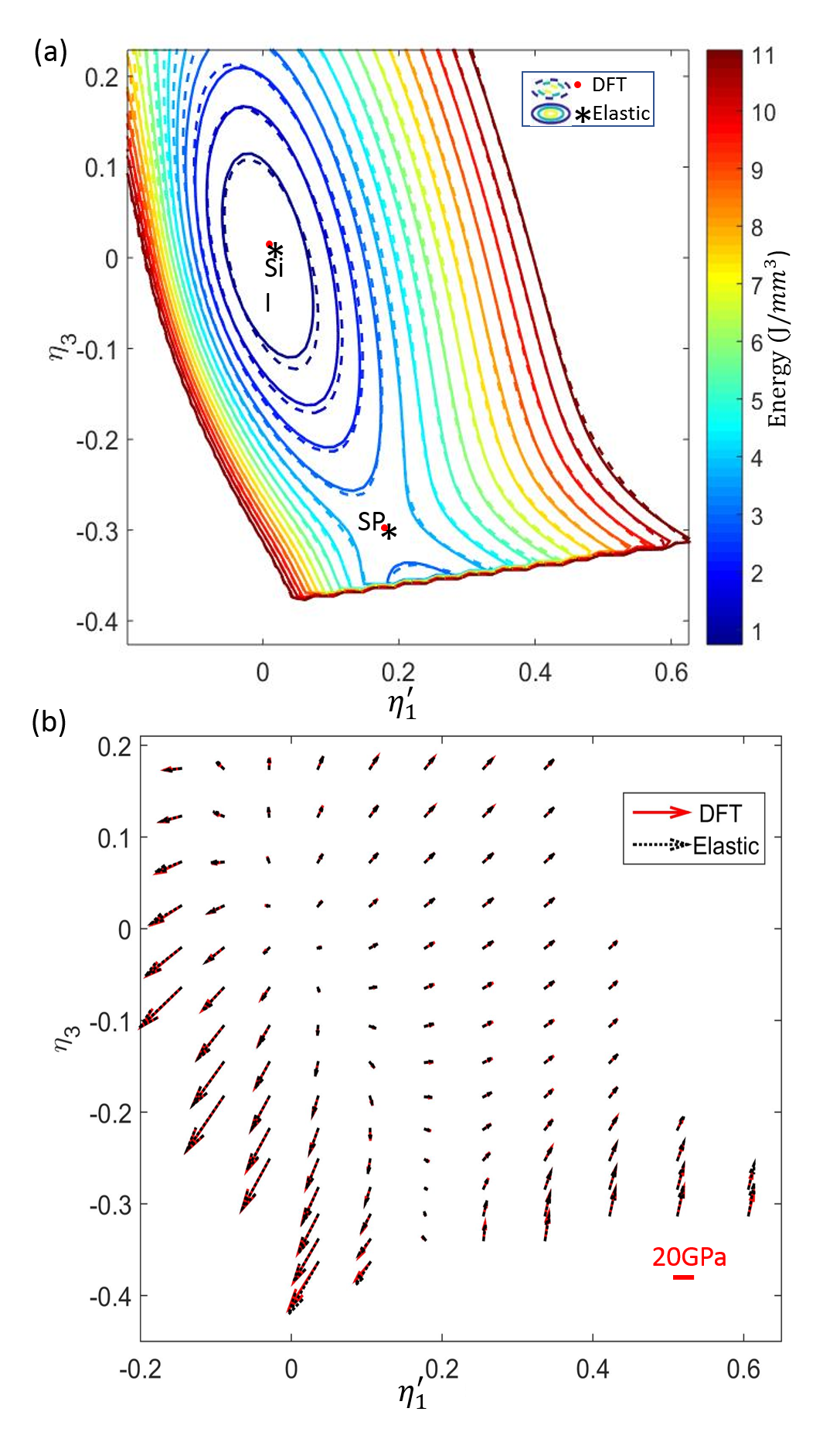}
    \caption{ For Si I, comparison  between analytical and DFT results for (a) fifth-order elastic energy and (b) energy gradients in $\eta'_1 =\eta'_2$ - $\eta_3$ plane. Components of gradients are PK2 stresses $S'_1=S'_2$ and $S_3$.}
    \label{fig:plot1}
\end{figure}

{\par}{\it Validation for energy. } Comparing energy contours from the elastic potential and DFT results in the plane of strains $\eta'_1 =\eta'_2$ and $\eta_3$ is given in Fig. \ref{fig:plot1}(a) ($\eta'_1 =\eta'_2$  are rotated by $45^0$ around axis 3 coordinate system, as in DFT unit cell \cite{supplement}).
The stress-free Si I from elastic approximation has lattice parameters $a_1=3.89\,$\AA, $c_1=5.47\, $\AA, within $1\%$ of DFT results ($a_1 = 3.8653\,$\AA, $c_1=5.4665 \,$\AA), and close to the recommended value of $5.431\, 020\, 511 (89) \,${\AA} \cite{CODATA2018}. The saddle point  (SP: $\eta'_1$ = $0.1777$ and $\eta_3$ = $-0.2584$)
has energy $3.2976$~J/mm$^{3}$ vs. $3.2893$~J/mm$^{3}$ from DFT.
The ability to yield the SP is crucial in capturing the elastic instabilities driving the phase transformation. Furthermore, in Fig. \ref{fig:plot1}(b), the  gradients of elastic energy in $\eta'_1 $-$\eta_3$ plane (with components equal to the PK2 stresses $S'_1=S'_2$ and $S_3$) from nonlinear elastic approximation correspond well to those from DFT. Deviations between the analytical results and DFT are quite small. Note that we did not aim to fit points far from the SP toward Si II as they should be fitted to the elastic potential for Si II.

\begin{figure} [!tbp]
 \centering
 \includegraphics[width=0.4\textwidth]{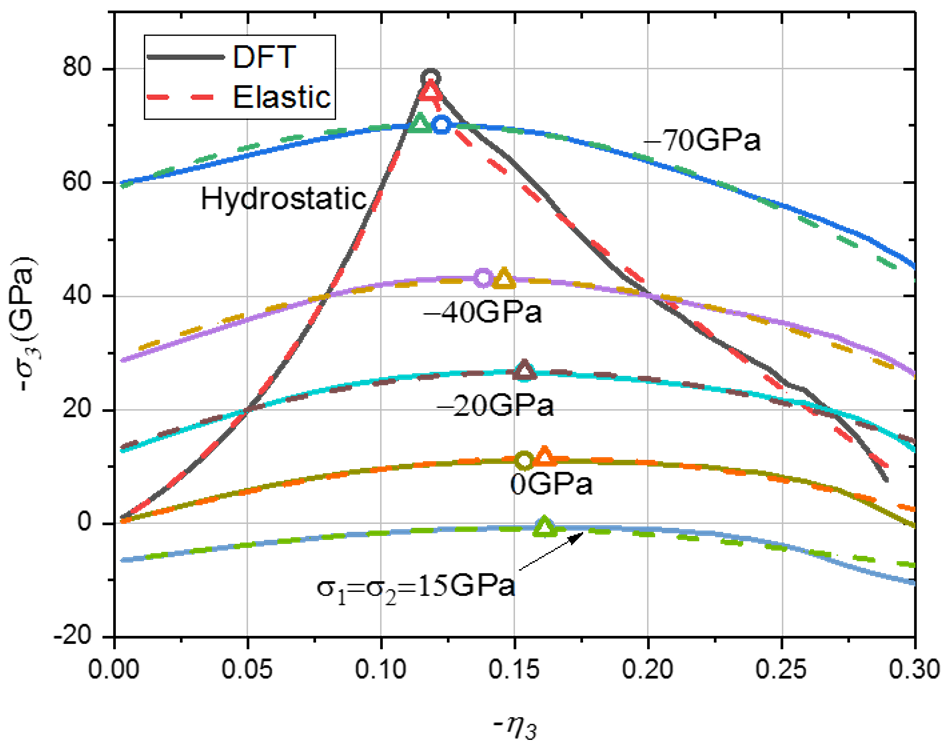}
 \caption{Cauchy (true) stress vs. Lagrangian strain ($\sigma_{3}$ vs. $\eta_{3}$) for $c$ axis compression or tension for different lateral stresses $\sigma_{1}=\sigma_{2}$ along Si I$\rightarrow$ Si II PT path.     DFT (circles) and elastic potential (triangles) designate results with maximum  $\sigma_{3}$.
 The excellent agreement between elastic potential and DFT is evident. }
    \label{fig:plot2}
\end{figure}

{\par}{\it Stress-strain curves for triaxial loading.} We compare  the Cauchy (true) stress $\sigma_{3}$ - $\eta_{3}$ curves  for different fixed lateral stresses $\sigma_{1}=\sigma_{2}$ along the path toward  Si  I$\rightarrow$ Si II PT (Fig. \ref{fig:plot2}).  Corresponding transformation paths in the ($\eta_1 = \eta_2, \eta_3$) plane are found iteratively using Newton method both for elastic potential and DFT simulations and are presented in \cite{supplement}.
It is clear from Fig. \ref{fig:plot2} that the fifth-order elastic potential captures the stress-strain curves from DFT calculations correctly for $0 \leq -\eta_{3} \leq 0.3$, including peak points of the stress-strain curves, corresponding to elastic instabilities. We  use the same definition  as in \cite{zarkevich2018lattice}: Elastic lattice instability at prescribed true stress $\sigma$ occurs at stresses above which the crystal cannot be at equilibrium.
All stress-strain curves are smooth, except one for hydrostatic loading. For nonhydrostatic loading, after instability point,
elastically distorted tetragonal lattice of Si I continues transformation to tetragonal Si II. However,  for
hydrostatic loading  a primary isotropic deformation of cubic Si I is getting unstable with respect to
 a secondary tetragonal perturbation leading to Si II. Such a bifurcation of the deformation path causes  discontinuity of the first derivative at the  instability point.
This bifurcation and jump in slope are  captured correctly in Fig. \ref{fig:plot2}.

{\it Elastic lattice instability criterion under triaxial loading.}
Combining lattice instability points from DFT and elastic potential, we present the lattice instability criterion in the form of the critical value $A$ of the modified transformation work:
\begin{equation} \label{eq:1}
    W = b_{3}\sigma_{3}\epsilon_{t3} + b_{1}(\sigma_1 + \sigma_2)  \epsilon_{t2}= A.
\end{equation}
Here $\vep_{t1}=\vep_{t2}=0.243$  and $\vep_{t3}=-0.514$  are transformation strain mapping stress-free crystal lattice of Si I into stress-free  lattice of Si II, and $b_{1}$ and $b_{3}$ are modifying constants. This criterion was derived in \cite{Levitas-IJP-13,Levitasetal-Instab-17,Levitasetal-PRL-17} via phase field and was verified and quantified by both molecular dynamics simulation using Tersoff potential \cite{Levitasetal-Instab-17} and DFT simulations \cite{zarkevich2018lattice}.   Instability lines  can be  approximated by  $\sigma_3 = 0.4144(\sigma_1+\sigma_2)-10.9121$ for nonlinear elasticity and by $\sigma_3 = 0.4066(\sigma_1+\sigma_2)-11.4493$ for DFT results, see  supplement \cite{supplement}.  Thus, our fifth-order elastic potential developed here successfully reproduces the lattice instability found in DFT over a range $0.5(\sigma_1+\sigma_2) \subset [-73.8; 16]$.
The strong effect of the nonhydrostatic stresses on the lattice instability is evident: the transformation pressure under hydrostatic loading is $\sim$ 75 GPa and transformation stress $\sg_3$ under uniaxial loading is $\sim$ 11 GPa (or mean stress of 3.7 GPa).

\begin{figure} [!tbp]
 \centering
 \includegraphics[width=0.4\textwidth]{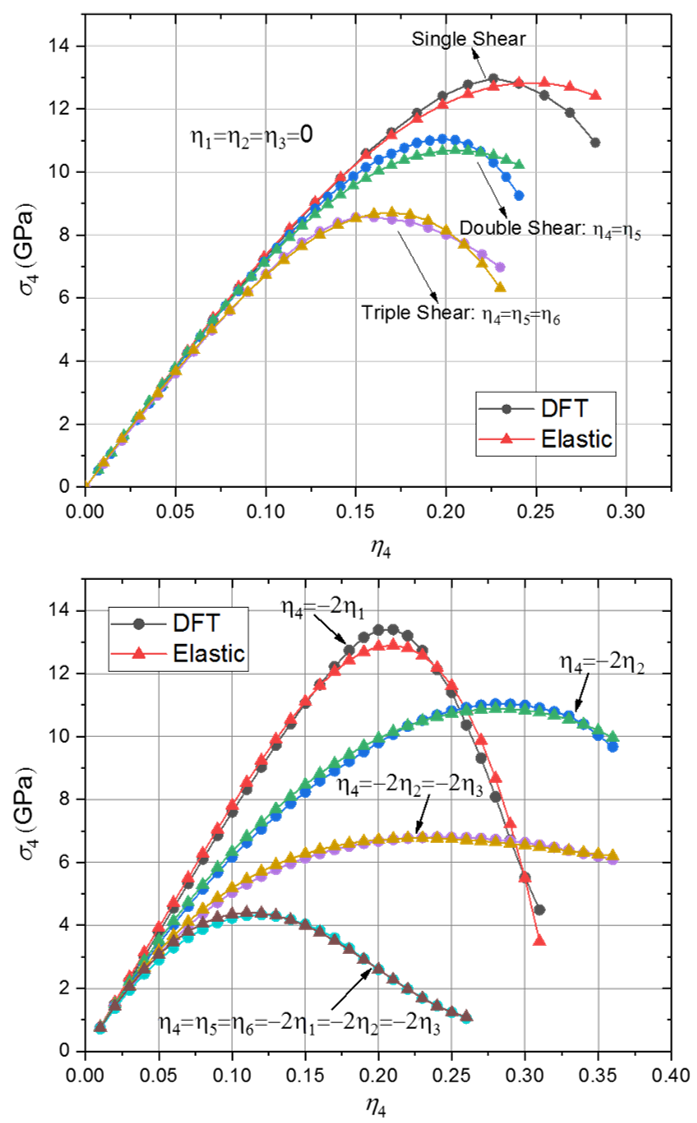}
 \caption{True shear stress--strain curves from 5$^{th}$-order elastic potential compared to DFT results: for (a) single, double, and triple simple shear strains ($\eta_{1}=\eta_{2}=\eta_{3}=0$); (b)  combination of normal and shear strains (all  non-mentioned strains are zero).
 Fifth-order potential describes DFT results well, including shear instabilities. }
    \label{fig:plot4}
\end{figure}

{\par} {\it Shear stress-strain curves \& instabilities under complex loading:}  Shear stress-strain curves  for simple shears (without normal strains) and for complex loading (shear plus normal strains) are shown in Fig. \ref{fig:plot4}.  The  elastic shear instability starts at $12.84$~GPa (DFT: $12.97$) for single shear, reduces to $10.7$~GPa (DFT: $11$) for double shear ($\eta_4=\eta_5$), and then to $8.71$~GPa (DFT: $8.56$) for triple shear, below $3\%$ error with DFT for strains beyond the instability points. Due to symmetry with respect to sign change, there are fewer nonzero elastic constants for shear than for normal strains; for single shear $\eta_4$, $c_{444}=c_{44444}=0$, and third and fifth degrees of $\eta_4$ and $\eta_5$ and $\eta_6$ are absent. Expectedly,  deviation of elastic approximation from DFT grows for strains beyond the shear instability points much faster than for normal  strains in Fig. \ref{fig:plot2}. This is not critical, as for unstable branch a phase transformation occurs, which is better described by the order parameter \cite{levitas2018phase,babaei2018phase}.
 Note, in a molecular dynamics simulation \cite{chen2019amorphization} with a Stillinger-Weber  potential \cite{stillinger1985computer}, the instability for simple shear along  $\left<\bar{11}2 \right>$  in the $\left(111 \right)$ plane and  along the $\left<111 \right>$ in the $\left(1\bar{1}0 \right)$ plane lead to amorphization.

 {\par} Note that double and triple shears along the $\left<100\right>$ in the $\left(001 \right)$ plane in Fig. \ref{fig:plot4}(a) represent single shear in $\left<110\right>$ in the $\left(001 \right)$ plane with $\eta_4^{'}= \sqrt{2} \eta_4$  and triaxial normal-strain loading in $\left<111\right>$ and in the $\left(111 \right)$ plane with $\eta_1^{'}=2\eta_4$ and $\eta_2^{'}=\eta_3^{'}=-\eta_4$, respectively.
 Then curves in Fig. \ref{fig:plot4}(a) can be analyzed in terms of the effect of  crystallographic anisotropy.
 Generally, by rotating coordinate system and transforming elastic potential accordingly, one can
 study the effect of the anisotropy for an arbitrary complex loading.

{\par} For the shearing in combination with  compressive normal strains (Fig. \ref{fig:plot4}(b)), the DFT results are described
by our  elastic potential even better than just for shearing, i.e., with smaller deviation for larger strains even exceeding 0.35. Interestingly, superposing uniaxial compression $\eta_1=-0.5 \eta_4$ orthogonal to shear plane  in Fig. \ref{fig:plot4}(b) slightly increases ultimate (theoretical) shear strength but slightly reduces corresponding shear strain
in comparison with Fig. \ref{fig:plot4}(a). At the same time, superposing uniaxial compression $\eta_2=-0.5 \eta_4$ in the  shear $\eta_4$ direction reduces ultimate shear strength by $\sim 2$~GPa, but increases corresponding shear strain. Superposing biaxial compression $\eta_1=\eta_2=-0.5 \eta_4$  further reduces ultimate shear strength down to $6.77$~GPa ($6.79$ from DFT) with corresponding shear strain between two previous cases. Shape of shear stress-strain curves changes also significantly with superposition of different compressive strains.  Also, superposing isotropic compression $\eta_1$ = $\eta_2$ = $\eta_3$ = $-0.5 \eta_4$ = $-0.5\eta_5$ = $-0.5 \eta_6 $ on the triple shearing   reduces  ultimate  shear strength from $8.71$~GPa ($8.56$ from DFT) in Fig. \ref{fig:plot4}(a)  to $4.4$~GPa ($4.31$ from DFT) in Fig. \ref{fig:plot4}(b) and also strongly reduces corresponding shear strain.
The tendency in reducing shear stability under hydrostatic loading in combination with presence of the dislocations
with local stress concentrators may lead to pressure-induced  amorphization observed experimentally  \cite{Debetal-Nat-01}. The observed coupling between shear and normal stresses is very nontrivial and well captured.  Typically, shear instabilities do not lead to Si II but rather to possible amorphization,  hexagonal diamond Si IV, slip, or twinning.

{\par} Note that presence of the plateau-like portion in the stress-strain curves for diamond was coined in \cite{liusmooth} as "atomic plasticity" and was considered as an indicator of desired combination of high strength with sufficient ductility. Such an atomic ductility is observed for Si under compression (Fig. \ref{fig:plot2} (a)) but not for shears (Fig. \ref{fig:plot4}(a)). However, superposition of certain normal strains (e.g., $\eta_2=-0.5 \eta_4$ and especially $\eta_1=\eta_2=-0.5 \eta_4$ ) significantly increases plateau.

{\it In summary,} the fifth-degree elastic potential for Si I under large strain including instability points was obtained in terms of Lagrangian strains by minimizing error relative to DFT results. Elastic energy and  true stress-strain curves for arbitrary complex loadings (including elastic instability) reproduce  DFT results very well.  Phase transition conditions for Si I$\rightarrow$Si II  under three normal
cubic  stresses are found to be linear in true stresses, in perfect agreement with DFT.  Any lower-order potentials (less than fifth-degree) cannot derive a similar precision in description of elastic instabilities and stress-strain curves, whereas, in contrast, they are currently found mostly using third-order elastic constants determined at small strains. Our results  also show the potential of controlling the stress-strain curves and phase transitions by applying optimized, multidimensional loading to control desirable properties and to drastically reduce phase transition pressures (1--2 orders of magnitude) \cite{zarkevich2018lattice,levitas2012virtual,Ji-etal-12,gao2019shear}.

{\par} Besides being generally applicable, the elastic  potential contains in convenient analytical form a plethora of information and now permits a direct study of all elastic instabilities   under complex loading driving different phase transitions (allotropic, amorphization, and melting), fracture, slip, and twinning.  Using higher-order potentials and large strains that include instabilities yields qualitatively and quantitatively better predictive capability, improving the entire model-based simulations, which are much faster than DFT.  Notably, our approach represents a fundamentally new basis for continuum simulations of crystal behavior under extreme static and dynamic loadings involving multiple the above mentioned orientational-dependent mechanisms.  In particular, higher-order elasticity is required for determination of the stress-strain states and optimization of the
diamond anvil cell for reaching maximum possible pressures  \cite{levitas2019tensorial}.  This approach is general and will significantly improve phase fields models of phase transformations, in contrast to second-order elasticity used currently \cite{levitas2018phase,babaei2018phase,Babaei-Levitas-PRL-20}.  It also provides a basis for the description
of the competition between different instabilities at different
loadings.

{\bf Acknowledgements:} VIL and HC are supported by NSF (CMMI-1536925 \& MMN-1904830),  ARO (W911NF-17-1-0225),  ONR (N00014-16-1-2079), \&
 XSEDE (MSS170015). NAZ and DDJ are supported by the U.S. Department of Energy (DOE), Office of Science, Basic Energy Sciences, Materials Science \& Engineering Division.
Ames Laboratory is operated for DOE by Iowa State University under contract DE-AC02-07CH11358.

\bibliography{Si-PRL}

\end{document}


\title{Fifth-degree elastic potential for predictive stress-strain relations and elastic instabilities under large strain and complex loading in Si}

\begin{center}
{\bf  SUPPLEMENTAL MATERIAL}
\end{center}

\author{Hao Chen}\email{chenhao.hitgucas@gmail.com}
\affiliation{Key Laboratory of Pressure Systems and Safety, Ministry of Education, School of Mechanical and Power Engineering, East China University of Science and Technology, Shanghai 200237, China}

\author{Nikolai~A. Zarkevich}\email{zarkev@ameslab.gov}
\affiliation{Ames Laboratory, U.S. Department of Energy, Iowa State University, Ames, Iowa 50011-3020, USA}

\author{Valery I. Levitas}\email{vlevitas@iastate.edu}
\affiliation{Ames Laboratory, U.S. Department of Energy, Iowa State University, Ames, Iowa 50011-3020, USA}
\affiliation{Department of Aerospace Engineering, Iowa State University, Ames, Iowa 50011, USA}
\affiliation{Department of Mechanical Engineering, Iowa State University, Ames, Iowa 50011, USA}
\affiliation{Department of Materials Science \& Engineering, Iowa State University, Ames, Iowa 50011, USA}

\author{Duane~D. Johnson}\email{ddj@iastate.edu, ddj@ameslab.gov}
\affiliation{Ames Laboratory, U.S. Department of Energy, Iowa State University, Ames, Iowa 50011-3020, USA}
\affiliation{Department of Materials Science \& Engineering, Iowa State University, Ames, Iowa 50011, USA}

\author{Xiancheng Zhang}
\affiliation{Key Laboratory of Pressure Systems and Safety, Ministry of Education, School of Mechanical and Power Engineering, East China University of Science and Technology, Shanghai 200237, China}

\maketitle


\section{\label{METHOD}SIMULATION METHODS}

We used DFT as implemented in VASP \cite{VASP1,VASP2,VASP3} with the projector augmented waves (PAW) basis \cite{PAW,PAW2} and PBE exchange-correlation functional \cite{PBE}.
The PAW-PBE pseudo-potential of Si had 4 valence electrons ($s^2 p^2$) and 1.9 {\AA} cutoff radius.
The plane-wave energy cutoff (ENCUT) was 306.7~eV, while the cut-off energy of the plane wave representation of the augmentation charges (ENAUG) was 322.1~eV.
We used a Davidson block iteration scheme (IALGO=38) for the electronic energy minimization.
Electronic structure was calculated with a fixed number of bands (NBANDS=16) in a tetragonal 4-atom unit cell (a supercell of a 2-atom primitive cell).
Brillouin zone integrations were done in $k$-space (LREAL=FALSE) using a $\Gamma$-centered Monkhorst-Pack mesh \cite{MonkhorstPack1976} containing 55 to 110 $k$-points per {\AA$^{-1}$} (fewer during atomic relaxation, more for the final energy calculation).
Accelerated convergence of the self-consistent calculations was achieved using a modified Broyden's method \cite{PRB38p12807y1988}.

{\par} Atomic relaxation in a fixed unit cell (ISIF=2) was performed using the conjugate gradient algorithm (IBRION=2), allowing symmetry breaking (ISYM=0).
The transformation path was confirmed by the nudged-elastic band (NEB) calculations, performed using the C2NEB code \cite{C2NEB}.
We used DFT forces in \emph{ab initio} molecular dynamics (MD) to verify stability of the relaxed atomic structures.
Si atoms were assumed to have mass POMASS=28.085 atomic mass units (amu).
The time step for the atomic motion was set to POTIM=0.5 fs.
Convergence vs. plane-wave energy cutoff is discussed in \cite{zarkevich2018lattice}.
The converged DFT data contained over $10^4$ entries and was processed in the format, outlined in Table 3 in \cite{Complexity11p36y2006}.

{\par}
The unit cells in the DFT simulations contained 4 atoms and were oriented along $A^{*} = \left \langle 110 \right \rangle$, $B^{*} = \left \langle 1\bar{1}0 \right \rangle$, and $C^{*} = \left \langle 001 \right \rangle$.
The transformation matrix for transforming the current simulation coordinate system to the natural $8$-atom cubic cell, oriented along $A = \left \langle 100 \right \rangle$, $B = \left \langle 010 \right \rangle$, and $C = \left \langle 001 \right \rangle$,  is:
\begin{equation} \label{eq:15}
R =[A, B, C][A^*, B^*, C^*]^{-1} = \begin{bmatrix}
1/\sqrt{2} & -1/\sqrt{2} & 0\\
1/\sqrt{2} & 1/\sqrt{2} & 0\\
0 & 0 & 1
\end{bmatrix}
\end{equation}
The transformation formulas of the deformation gradient and Cauchy and second Piola-Kirchhoff (PK2) stresses to the natural cubic coordinate system are
\begin{equation} \label{eq:16}
    F = R*F^{*}*R^{T}; \quad
    \sigma = R*\sigma^{*}*R^{T}; \quad  S = R*S^{*}*R^{T}.
\end{equation}
Parameter identification procedure is carried out and results are presented in the natural  cubic  coordinate system.

\section{\label{SSCURVES}Fifth-order elastic energy formula for cubic systems} 
The fifth-order energy for cubic symmetry can be expressed as
\begin{equation} \label{eq:10}
   u =\psi_{2}+\psi_{3}+\psi_{4}+\psi_{5}  + \cdots.
\end{equation}
in which
\begin{equation} \label{eq:11}
\begin{aligned}
    \psi_{2} =\frac{1}{2}c_{11}(\eta_{1}^2+\eta_{2}^2+\eta_{3}^2)+
    c_{12}(\eta_{1}\eta_{2}+\eta_{2}\eta_{3}+\eta_{1}\eta_{3})+
    \frac{1}{2}c_{44}(\eta_{4}^2+\eta_{5}^2+\eta_{6}^2)
\end{aligned}
\end{equation}
\begin{equation} \label{eq:12}
\begin{aligned}
    \psi_{3} =\frac{1}{6}c_{111}(\eta_{1}^3+\eta_{2}^3+\eta_{3}^3)+
    \frac{1}{2}c_{112}[\eta_{1}^2(\eta_{2}+\eta_{3})+
    \eta_{2}^2(\eta_{1}+\eta_{3})+
    \eta_{3}^2(\eta_{1}+\eta_{2})]+ \\
    \frac{1}{2}c_{155}[\eta_{4}^2(\eta_{2}+\eta_{3})+
             \eta_{5}^2(\eta_{1}+\eta_{3})+
             \eta_{6}^2(\eta_{1}+\eta_{2})] +
             c_{456}\eta_{4}\eta_{5}\eta_{6} +
    c_{123}\eta_{1}\eta_{2}\eta_{3}+ \\
    \frac{1}{2}c_{144}(\eta_{1}\eta_{4}^2+
            \eta_{2}\eta_{5}^2+
            \eta_{3}\eta_{6}^2)
\end{aligned}
\end{equation}
\begin{equation} \label{eq:13}
\begin{aligned}
    \psi_{4} = \frac{1}{4}c_{1122}(\eta_{1}^2\eta_{2}^2+
       \eta_{2}^2\eta_{3}^2+
       \eta_{1}^2\eta_{3}^2)+
    \frac{1}{6}c_{1112}[\eta_{1}^3(\eta_{2}+\eta_{3})+
  \eta_{2}^3(\eta_{1}+\eta_{3})+
  \eta_{3}^3(\eta_{1}+\eta_{2})]+ \\
       \frac{1}{24}c_{1111}(\eta_{1}^4+\eta_{2}^4+\eta_{3}^4)+
    \frac{1}{2}c_{1123}\eta_{1}\eta_{2}\eta_{3}(\eta_{1}+\eta_{2}+\eta_{3})+ \frac{1}{4}c_{1144}(\eta_{1}^2\eta_{4}^2+
                    \eta_{2}^2\eta_{5}^2+
                    \eta_{3}^2\eta_{6}^2)+ \\
    \frac{1}{4}c_{1155}[\eta_{1}^2(\eta_{5}^2+\eta_{6}^2)+
    \eta_{2}^2(\eta_{6}^2+\eta_{4}^2)+
    \eta_{3}^2(\eta_{5}^2+\eta_{4}^2)]+ \frac{1}{2}c_{1266}(\eta_{1}\eta_{2}\eta_{6}^2 +    \eta_{2}\eta_{3}\eta_{4}^2+
    \eta_{1}\eta_{3}\eta_{5}^2)\\
    \frac{1}{2}c_{1255}[\eta_{1}\eta_{2}(\eta_{4}^2+\eta_{5}^2)
    +\eta_{2}\eta_{3}(\eta_{5}^2+\eta_{6}^2)+
    \eta_{1}\eta_{3}(\eta_{4}^2+\eta_{6}^2)]+
    \frac{1}{4}c_{4455}(\eta_{4}^2\eta_{5}^2+\eta_{5}^2\eta_{6}^2+\eta_{4}^2\eta_{6}^2) + \\
    c_{1456}\eta_{4}\eta_{5}\eta_{6}(\eta_{1}+\eta_{2}+\eta_{3})+
    \frac{1}{24}c_{4444}(\eta_{4}^4+\eta_{5}^4+\eta_{6}^4)+
    \frac{1}{2}c_{1266}(\eta_{1}\eta_{2}\eta_{6}^2+
    \eta_{2}\eta_{3}\eta_{4}^2+
    \eta_{1}\eta_{3}\eta_{5}^2)
\end{aligned}
\end{equation}
\begin{equation} \label{eq:14}
\begin{aligned}
    \psi_{5} = \frac{1}{120}c_{11111}(\eta_{1}^5+\eta_{2}^5+\eta_{3}^5) + \frac{1}{2}c_{11456}\eta_{4}\eta_{5}\eta_{6}(\eta_{1}^2+\eta_{2}^2+\eta_{3}^2)+
    \frac{1}{2}c_{12344}\eta_{1}\eta_{2}\eta_{3}(\eta_{4}^2+\eta_{5}^2+\eta_{6}^2)+ \\
    \frac{1}{24}c_{15555}[\eta_{1}(\eta_{5}^4+\eta_{6}^4)+
           \eta_{2}(\eta_{4}^4+\eta_{6}^4)+
           \eta_{3}(\eta_{4}^4+\eta_{5}^4)]+
    \frac{1}{4}c_{11223}(\eta_{1}\eta_{2}^2\eta_{3}^2+
                   \eta_{1}^2\eta_{2}\eta_{3}^2+
                   \eta_{1}^2\eta_{2}^2\eta_{3})\\
    \frac{1}{24}c_{14444}(\eta_{1}\eta_{4}^4+\eta_{2}\eta_{5}^4+\eta_{3}\eta_{6}^4)+
    \frac{1}{4}c_{14455}[\eta_{1}\eta_{4}^2(\eta_{5}^2+\eta_{6}^2)+
            \eta_{2}\eta_{5}^2(\eta_{4}^2+\eta_{6}^2)+
            \eta_{3}\eta_{6}^2(\eta_{4}^2+\eta_{5}^2)]+ \\
    \frac{1}{12}c_{11122}[\eta_{1}^3(\eta_{2}^2+\eta_{3}^2)+
    \eta_{2}^3(\eta_{1}^2+\eta_{3}^2)+
    \eta_{3}^3(\eta_{1}^2+\eta_{2}^2)]+
    \frac{1}{6}c_{11123}(\eta_{1}^3\eta_{2}\eta_{3}+
    \eta_{1}\eta_{2}^3\eta_{3}+
    \eta_{1}\eta_{2}\eta_{3}^3)+ \\
    \frac{1}{12}c_{11155}[\eta_{1}^3(\eta_{5}^2 + \eta_{6}^2) +
                         \eta_{2}^3(\eta_{4}^2 + \eta_{6}^2) +
                         \eta_{3}^3(\eta_{4}^2 + \eta_{5}^2)] +
    \frac{1}{12}c_{11144}(\eta_{1}^3\eta_{4}^2 + \eta_{2}^3\eta_{5}^2 +
                       \eta_{3}^3\eta_{6}^2) +  \\
  c_{12456}\eta_{4}\eta_{5}\eta_{6}(\eta_{1}\eta_{2}+\eta_{1}\eta_{3}+\eta_{2}\eta_{3}) +
    \frac{1}{24}c_{11112}[\eta_{1}^4(\eta_{2}+\eta_{3})+
    \eta_{2}^4(\eta_{1}+\eta_{3})+
    \eta_{3}^4(\eta_{1}+\eta_{2})]+\\
     \frac{1}{4}c_{11266}[\eta_{1}\eta_{2}(\eta_{2}+\eta_{1})\eta_{6}^2+
        \eta_{1}\eta_{3}(\eta_{1}+\eta_{3})\eta_{5}^2+
        \eta_{2}\eta_{3}(\eta_{2}+\eta_{3})\eta_{4}^2]+ \\
        \frac{1}{4}c_{15566}(\eta_{3}\eta_{4}^2\eta_{5}^2+
                 \eta_{2}\eta_{4}^2\eta_{6}^2+
                 \eta_{1}\eta_{5}^2\eta_{6}^2)+
\frac{1}{6}c_{44456}\eta_{4}\eta_{5}\eta_{6}(\eta_{4}^2+\eta_{5}^2+\eta_{6}^2) + \\
\frac{1}{4}c_{12244}[\eta_{1}(\eta_{2}^2+\eta_{3}^2)\eta_{4}^2+
           \eta_{2}(\eta_{1}^2+\eta_{3}^2)\eta_{5}^2+
           \eta_{3}(\eta_{1}^2+\eta_{2}^2)\eta_{6}^2]+ \\
\frac{1}{4}c_{11244}[\eta_{1}^2(\eta_{2}+\eta_{3})\eta_{4}^2+
                \eta_{2}^2(\eta_{1}+\eta_{3})\eta_{5}^2+
                \eta_{3}^2(\eta_{2}+\eta_{1})\eta_{6}^2]
\end{aligned}
\end{equation}

\newpage
\section{\label{add-res}Some additional results} 

\begin{figure} [!tbp]
 \centering
 \includegraphics[width=0.45\textwidth]{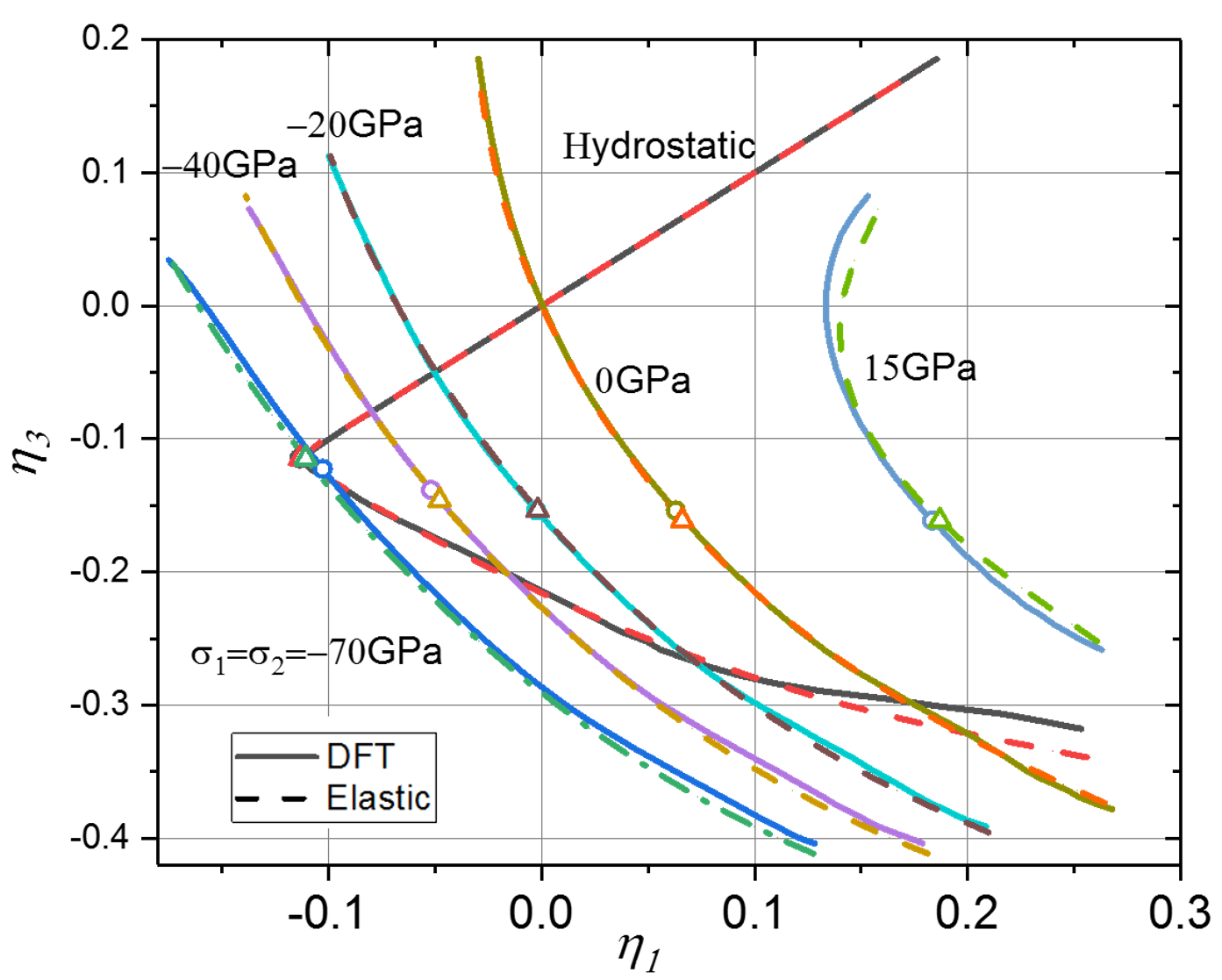}
  \caption{Comparison of the loading paths obtained using DFT simulations and elastic potential in the ($\eta_1 =\eta_2, \eta_3$) plane for compression/tension along $\eta_{3}$ for different fixed stresses  $\sigma_{1}=\sigma_{2}$ corresponding to the results in Fig. 2 of the main text.
  The  "hydrostatic" line designates loading path for $\sigma_{1}=\sigma_{2}=\sigma_3$.  }
    \label{fig:plot20}
\end{figure}

Comparison of the loading paths corresponding to  compression or tension  along $\eta_{3}$ at different fixed stresses  $\sigma_{1}=\sigma_{2}$  from nonlinear elastic approximation and DFT results is presented in Fig.~\ref{fig:plot20}.
Good correspondence is evident. These paths are used  to obtain  stress $\sigma_{3}$ vs. Lagrangian strain $\eta_{3}$ dependence in Fig. 2 of the main text.

\begin{figure} [!tbp]
 \centering
 \includegraphics[width=0.45\textwidth]{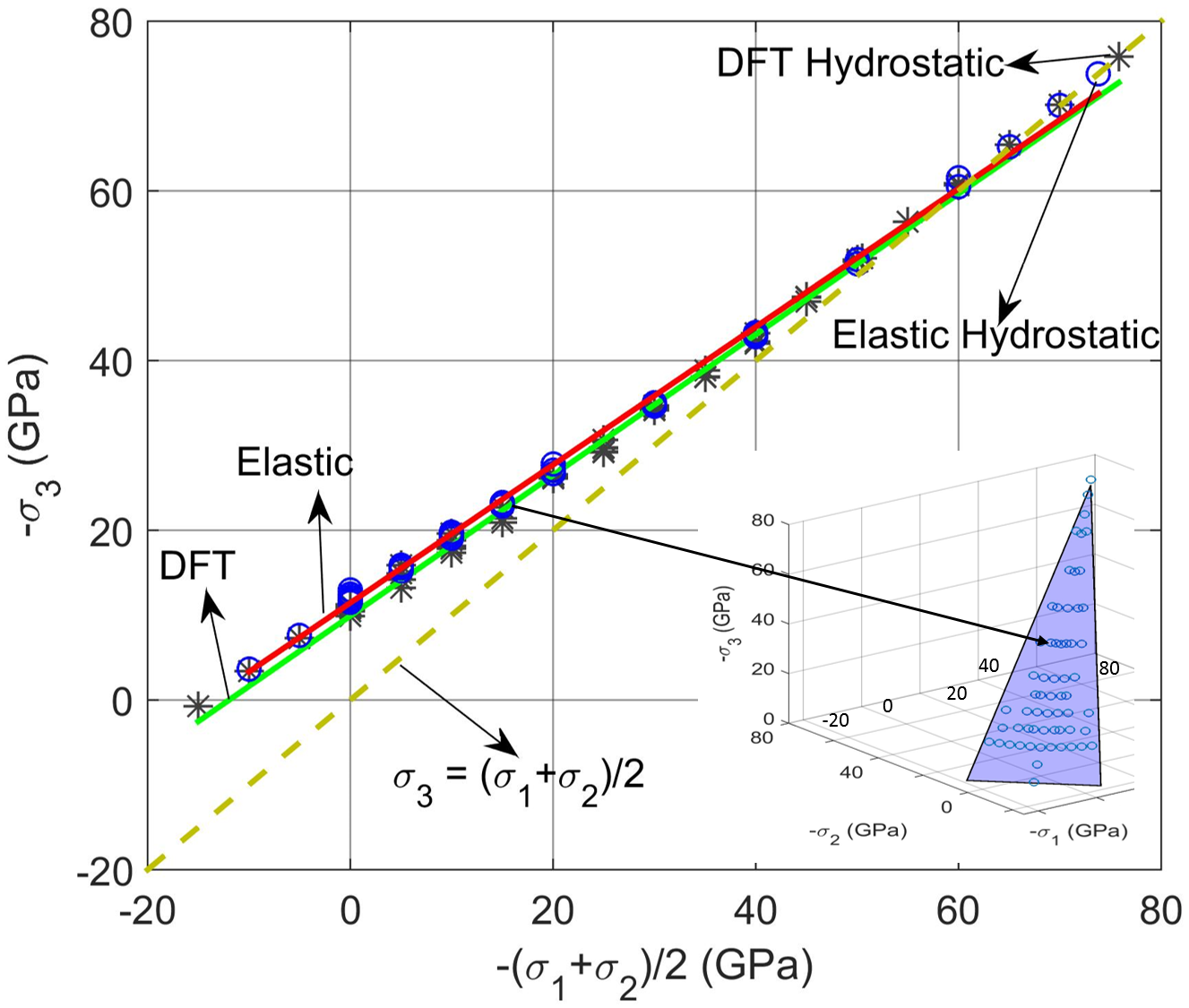}
 \caption{Comparison of the elastic instability condition $\sigma_3$ versus $0.5(\sigma_1 + \sigma_2)$ for  Si I $\rightarrow$ Si II PT from the fifth-order elastic potential (circles) and DFT results (*). The inset  shows the same instability points from the fifth-order elastic potential  (circles)    in $3D$ space $\sigma_{1}$, $\sigma_{2}$, and $\sigma_{3}$, which  lie very close to the instability plane calibrated with DFT in \cite{zarkevich2018lattice}.}
    \label{fig:plot10}
\end{figure}

Lattice instability lines under triaxial compression stress states for  PT Si I $\rightarrow$ Si II are presented in Fig. \ref{fig:plot10}. They are approximated by linear function in the main text. Again, they practically coincide.

\newpage
\bibliography{Si-PRL}